\documentclass[12pt]{article}
\usepackage[dvips]{epsfig}

\def\be{\begin{equation}}
\def\eq{\end{equation}}

\def\q2{{\rm QCD}_{2}}
\def\q4{{\rm QCD}_{4}}

\begin{document}
\begin{flushright}
hep-th/9707146
\end{flushright}
\vspace{20mm}
\begin{center}
{\LARGE  On Matrix Strings, the Large $N$ Limit and \\
 Discretized Light-Cone Quantization \\}
\vspace{20mm}
{\bf ${}^{*,\#}$F. Antonuccio\footnote{
Address from August 1997: The Ohio State University, Columbus OH, USA}
, ${}^{*}$H.C. Pauli,
${}^{*}$S.Tsujimaru\\}
\vspace{10mm}
{\em ${}^{*}$Max-Planck-Institut f\"{u}r Kernphysik, 69029 Heidelberg, Germany
\\ and \\
${}^{\#}$Institute for Theoretical Physics, University of Heidelberg,
69120 Heidelberg, Germany}
\end{center}
\vspace{20mm}
\begin{abstract}
We consider the $1+1$ dimensional supersymmetric matrix field
theory obtained from a dimensional reduction of ten dimensional 
${\cal N} = 1$  super Yang-Mills, which is a matrix model candidate
for non-perturbative Type IIA string theory. The gauge group here 
is U($N$), where $N$ is sent to infinity. 
We adopt light-cone coordinates to parametrize the string
world sheet, and choose to work in the light-cone gauge.
Quantizing this theory via Discretized Light-Cone Quantization (DLCQ) 
introduces an integer, $K$, which restricts
the light-cone momentum-fraction of constituent quanta to be
 integer multiples of
$1/K$.  
We show
how a double scaling limit involving
the integers $K$ and $N$ implies the existence of 
an extra (free) parameter in the Yang-Mills theory, which
plays the role of an effective string coupling constant.
The formulation here provides a natural framework for
studying quantitatively string dynamics and 
conventional Yang-Mills in a unified setting.

\end{abstract}
\newpage

\baselineskip .25in

\section{Introduction}
Much of the recent excitement in string theory 
stems from a conjecture
about the essential degrees of freedom underlying $M$-theory;
namely, eleven dimensional $M$-theory is given
by a U($N$) gauge invariant
supersymmetric matrix model, 
which is formally obtained
by reducing $9+1$ dimensional 
super Yang-Mills to $0+1$ dimensions via classical dimensional
reduction \cite{shenk}.  
Motivated by the work of Witten \cite{witten1},
we may interpret the eigenvalues of the (Hermitian matrix-valued)
Yang-Mills fields as space-time coordinates of D-particles,
where the world line trajectories are given by the $0+1$ dimensional
matrix model Lagrangian. Deviations from classical space-time are therefore
seen from the non-commutative properties of these matrix coordinates.
Of course, until the correspondence between $M$-Theory
and the matrix model is
rigorously established, we have to content ourselves with the
label ``Matrix Theory'' for the matrix model, 
in order to distinguish it (and
its consequences) from 
the formal definition of $M$-theory \cite{banks}. 
   
Now the underlying Yang-Mills theory of the matrix model is
manifestly ten dimensional, and so in order to establish 
any connection with (eleven dimensional) $M$-Theory, 
a necessary (and highly non-trivial) ingredient in the conjecture
is the assertion that the large $N$ limit  
of the matrix model
effectively gives rise to an additional space-like dimension.
The resulting theory is then interpreted as $M$-Theory in
the light-cone frame. An attempt to strengthen the
plausibility of this assumption by associating the integer $N$
with the harmonic resolution $K$ of
Discretized Light-Cone Quantization (DLCQ)\cite{pauli}
has recently been made by Susskind\cite{suss1}.

Evidently, these developments suggest that another 
(closer) look at 
Yang-Mills theory in the large $N$ limit
is in order if we wish to understand its true 
underlying dynamics, and, ironically, obtain possible insights into 
a new kind of `string' theory. The crucial issue here
is the (conjectured) existence of a  
{\em double scaling limit} involving the 
integer $N$, which we send 
to infinity,
and an additional parameter -- such as
a lattice spacing, $\epsilon$ --  which must eventually
be sent to zero \cite{kawai}. 
Simply put: whether one takes these limits independently or not 
determines whether one ends up with conventional
Yang-Mills, or not. The result is that we have an extra
free parameter, in addition to
the usual coupling constant, and so we are forced to
generalize our usual concept of Yang-Mills theory.
Moreover, this new parameter
can be shown to be related to the 
string coupling constant of Type IIA or
Type IIB  string theory \cite{kawai,witten2}. 
Conventional Yang-Mills is then recovered when this
extra parameter is set to zero. 

In summary, developments in non-perturbative
string theory have shed new light on the interpretation
of large $N$ Yang-Mills theory, giving rise to
`generalized Yang-Mills'\cite{witten2}. We will discuss in this article
how DLCQ enables one to clarify these ideas in a 
quantitative manner.   

\begin{figure}[h!]
\begin{center}
\epsfig{file=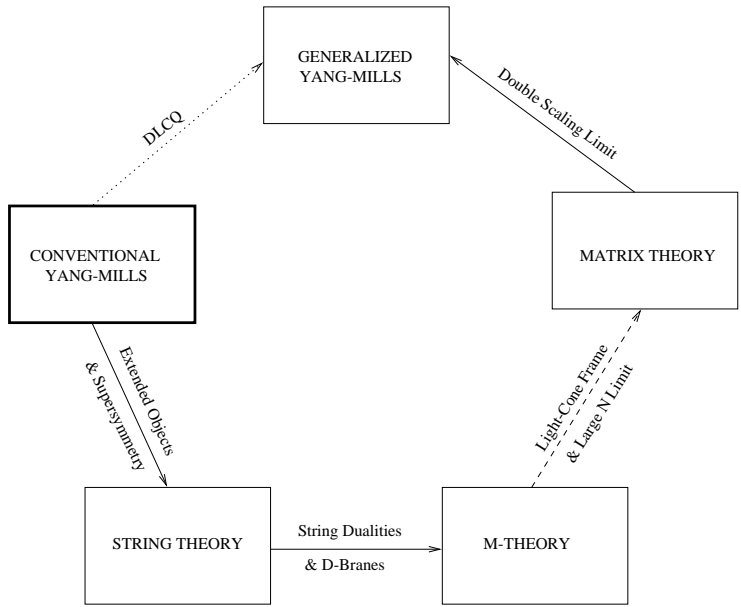, height=8cm}
\end{center}
\caption{{\small The emergence of `generalized' Yang-Mills.} 
\label{pentagon}}
\end{figure}

Figure \ref{pentagon} outlines how a notion of generalized 
Yang-Mills may be viewed as
emerging from string theory related 
investigations\footnote{Evidently, space limitations prevent a more detailed
representation than the one given here.};
the arrows are labeled by key concepts involved in the 
progress from one development to the next, starting with
ordinary Yang-Mills, and moving anti-clockwise around the figure
(the broken arrow relating formal $M$-Theory to Matrix Theory
reflects the conjectured equivalence between the two).
The connection between ordinary and generalized Yang-Mills
via DLCQ is also 
implied, and will be the subject
of this paper.

Our main goal therefore is to outline a framework which 
(we believe) naturally incorporates
string dynamics and conventional Yang-Mills in
a unified setting. 
The contents of this paper are organized as follows;
 in Section \ref{yang} we consider 
 the two dimensional super Yang-Mills
theory which is a candidate for non-perturbative Type IIA 
string theory 
\cite{motl,seiberg,verlinde}. We formulate the theory
using light-cone coordinates for the
string world sheet, and quantize the theory in the light-cone gauge.
In  Section \ref{dlcq}, we discuss 
Discretized Light-Cone Quantization.
This discretization procedure admits a natural
formulation of `bit strings', which
will represent our string states, and which  can be shown to
end on D-particle-like configurations. This is the subject matter of 
Section \ref{branes}.
Subtleties associated with the large $N$ limit, and the 
implications of an additional dimensionless
parameter in Yang-Mills due to a double scaling
limit will be addressed in Section \ref{doublescaling}.  
We conclude with a comment on possible future directions in 
Section \ref{end}.

\section{Matrix String Theory on the Light-Cone}
\label{yang}
By definition, compactifying eleven dimensional $M$-theory 
on a circle $S^1$ gives Type IIA string theory\footnote{
The radius of this 
circle
is identified with the string coupling constant, and so
$M$-theory is just the strong coupling limit of
ten dimensional Type IIA string theory.}.  
This suggests how one may obtain a possible representation
of non-perturbative
Type IIA string theory via Matrix Theory; namely, we 
compactify one of the spatial dimensions of the matrix 
model on a circle $S^1$ \cite{shenk,wati}. The theory
we end up with is called Matrix String Theory, although there
is now considerable evidence that it exhibits
the known properties of 
Type IIA string theory \cite{motl,seiberg,verlinde}. 

Actually, it turns out that Matrix String Theory may be formally obtained 
by dimensionally reducing $9+1$ dimensional ${\cal N}=1$
super Yang-Mills to $1+1$ dimensions. For the sake of completeness,
we review the underlying ten dimensional light-cone Yang-Mills theory
in Appendix \ref{ymills10} -- in perhaps more detail
than is customary -- although the ideas should be 
familiar to many
readers.  

In order to dimensionally reduce the ten dimensional Yang-Mills
action (\ref{LCversion}) (see Appendix \ref{ymills10})
to a $1+1$ dimensional field theory, we
simply specify that all fields are independent of the (eight) 
transverse
coordinates $x^I$, $I=1,\dots,8$ (points in 
Minkowski space are specified by the ten space-time
coordinates $(x^0,x^1,\dots,x^9)$).  We may therefore assume
that the fields depend only on the light-cone variables $\sigma^{\pm}
= \frac{1}{\sqrt{2}}(x^0 \pm x^9)$. The resulting two dimensional
theory has ${\cal N} = 8$ supersymmetry, and may be described by the action
\begin{eqnarray}
S_{1+1}^{LC} & = & \int d\sigma^+ d\sigma^- \hspace{1mm}
 \mbox{tr} \left( \frac{1}{2}D_\alpha X_I D^\alpha X_I + \frac{g^2}{4}
            [X_I,X_J]^2 - \frac{1}{4} F_{\alpha \beta} F^{\alpha \beta} 
 \right. \nonumber \\
& & \hspace{20mm}
+ \hspace{1mm}
{\rm i} \theta_R^T D_+ \theta_R +   {\rm i}\theta_L^T D_- \theta_L 
    - \sqrt{2}g\theta_L^T \gamma^I[X_I,\theta_R] \left. \frac{}{} \right),  
\label{LCversionreduced}
\end{eqnarray}
where the repeated indices $\alpha,\beta$ are summed over light-cone
labels $\pm$, and $I,J$ are summed over $1,\dots,8$.
The eight scalar fields $X_I(\sigma^+,\sigma^-)$ represent
$N \times N$ Hermitian matrix-valued fields, and are 
remnants of
the transverse components of the
ten dimensional gauge field $A_\mu$, while $A_{\pm}(\sigma^+,\sigma^-)$ 
are the
light-cone gauge field components of the residual 
two dimensional U($N$) gauge symmetry. 
The spinors $\theta_R$ and
$\theta_L$ are the remnants of
the right-moving and left-moving
projections of a sixteen component real spinor in the ten dimensional
theory. The components of  $\theta_R$ and $\theta_L$
are also  $N \times N$ Hermitian matrix-valued fields.  
  $F_{\alpha \beta} =  
\partial_{\alpha} A_\beta - \partial_\beta A_\alpha
    +{\rm i}g[A_\alpha, A_\beta]$ is just the two dimensional
gauge field tensor, while
$D_\alpha =  \partial_\alpha + {\rm i}g[A_\alpha,\cdot]$ is the covariant
derivative corresponding to the adjoint representation of the
gauge group U($N$). The eight $16 \times 16$ real symmetric matrices
$\gamma^I$ are defined in Appendix \ref{ymills10}.
 
In order to make a connection with string theory, we identify
the $1+1$ dimensional space parametrized by the light-cone coordinates
$(\sigma^+,\sigma^- )$ as the string world sheet. The eigenvalues
of the matrices $X_I$ are then identified with the target space
coordinates
of the Type IIA fundamental string \cite{verlinde}. Of course,
the $X_I$'s are non-commuting in general, and so we cannot simultaneously
diagonalize them to obtain a classical description of a 
propagating string. It is in this sense that Matrix (String) Theory
forces us to revise our notion of space-time as an approximately
derived concept, and deviations from the classical formulation
may be measured in terms of the non-commuting properties of these
matrix coordinates.

Since we are working in the light-cone frame, it is natural
to adopt the light-cone gauge $A_- = 0$. With
this gauge choice, the action (\ref{LCversionreduced}) becomes   
\begin{eqnarray}
{\tilde S}_{1+1}^{LC}&=&
\int d\sigma^+d\sigma^- {\rm {tr}} \Bigg(\partial_+X_I\partial_-X_I +
{\rm i} 
\theta_R^T\partial_+ \theta_R + {\rm i}\theta_L^T\partial_- \theta_L 
\nonumber\\
&+&\frac{1}{2}(\partial_-A_+)^2 +gA_+J^+ 
-\sqrt{2}g \theta_L^T \gamma^I [X_I, \theta_R ] 
+\frac{g^2}{4}[X_I, X_J ]^2\Bigg), 
\label{EQ6}
\end{eqnarray}
where $J^+ ={\rm i}[X_I, \partial_-X_I]+2\theta_R^T\theta_R$ 
is the longitudinal
momentum current.
The (Euler-Lagrange) equations of motion for the $A_+$
and $\theta_L$ fields are now
 \begin{eqnarray}
&&\partial_-^2A_+=gJ^+, \label{firstc}\\
&& \sqrt2 {\rm i}\partial_-\theta_L=g\gamma^I [X_I,\theta_R].
\label{secondc}  
\end{eqnarray}
These are evidently constraint equations, since
they are independent of the light-cone time $\sigma^+$.
The ``zero mode'' of the constraints above provide
us with the conditions
\begin{equation}
\int d\sigma^- J^+=0, \mbox{     and      } 
\int d\sigma^- \gamma^I [X_I,\theta_R] =0,
\label{EQ4}
\end{equation}    
which will be imposed on the Fock space 
to select the physical states in the quantum theory. 
The first constraint above is well known in the literature,
and projects out the colourless states in
the quantized theory\cite{kd1}. The second (fermionic) constraint is
perhaps lesser well known, but certainly provides  non-trivial
relations governing the small-$x$ behaviour of light-cone 
wavefunctions\footnote{If we introduce a mass term, such
relations become crucial in establishing finiteness 
conditions. See \cite{abd}, for example.} \cite{abd}. 
  
At any rate, equations (\ref{firstc}),(\ref{secondc}) permit one
to eliminate the non-dynamical fields $A_+$ and $\theta_L$  
in the theory, which is a particular feature of
light-cone gauge theories. There are no ghosts.
We may therefore write down explicit  expressions for the
light-cone momentum $P^+$ and Hamiltonian $P^-$ in terms
of the physical degrees of freedom of the theory, which
are denoted by the eight 
scalars $X_I$, and  right-moving spinor $\theta_R$: 
\begin{eqnarray}
P^+&=&\int d\sigma^- \hspace{1mm}
\mbox{tr}
\left( \partial_-X_I\partial_-X_I+{\rm i}
\theta_R^T \partial_-\theta_R \right), \label{P+}
\\  
P^- &=&g^2 
\int d\sigma^- {\rm {tr}}\Bigg(-\frac{1}{2} J^+\frac{1}{\partial_-^2}J^+
-\frac{1}{4}[X_I, X_J ]^2 \nonumber \\ 
&&\hspace{15mm}+\frac{{\rm i}}{2} 
(\gamma^I [X_I, \theta_R])^T
\frac{1}{\partial_-} \gamma^J [X_J, \theta_R]\Bigg). 
\label{P-}
\end{eqnarray}
The light-cone Hamiltonian propagates a given field configuration
in light-cone time $\sigma^+$, and contains all the non-trivial
dynamics of the interacting field theory. 

In the representation for the $\gamma^I$ matrices specified by 
(\ref{gamma9}) in Appendix \ref{ymills10}, we may write
\begin{equation}
\theta_R = { u \choose 0}, \label{spin8} 
\end{equation}
where $u$ is an eight component real spinor.      
The commutation relations at equal 
light-cone time $\sigma^+=\rho^+$ take the following 
form for $I,J,\alpha,\beta=1,\dots,8$:  
\begin{eqnarray}
&&[X^I_{pq}(\sigma^+,\sigma^-), \partial_-X^J_{rs}(\rho^+,\rho^-)]
=\frac{{\rm i}}{2}
\delta(\sigma^--\rho^-)\delta^{IJ}\delta_{ps}\delta_{qr},  \\
&&\{u_{pq}^{\alpha}(\sigma^+,\sigma^-), 
u^{\beta}_{rs}(\rho^+,\rho^-)\}=\frac{1}{2}
\delta(\sigma^--\rho^-)\delta^{\alpha \beta}\delta_{ps}\delta_{qr},   
\end{eqnarray}
where the lower indices of the fields  label the 
components of an $N\times N$ (Hermitian) matrix. 
In terms of their Fourier modes, the fields 
may be expanded at light-cone time $\sigma^+=0$ to give\footnote{
The symbol $\dagger$ denotes quantum conjugation, and does not
transpose matrix indices.} 
\begin{eqnarray}
&&X^I_{pq}(\sigma^-)= \frac{1}{\sqrt{2\pi}}
\int_{0}^{\infty}\frac{dk^+}{\sqrt{2 k^+}}\Big(a^I_{pq}(k^+)
e^{-{\rm i}k^+\sigma^-}
+ {a^I_{qp}}^{\dagger}(k^+)e^{{\rm i}k^+\sigma^-}\Big), 
\hspace{4mm} I=1,\dots,8; \hspace{3mm} \label{Xexp}\\
&&u^{\alpha}_{pq}(\sigma^-)=\frac{1}{\sqrt{2 \pi}}\int_0^{\infty}
\frac{dk^+}{\sqrt{2}} 
\Big(b^{\alpha}_{pq}(k^+)e^{-{\rm i}k^+\sigma^-}
+ {b^\alpha_{qp}}^{\dagger}(k^+)e^{{\rm i}k^+\sigma^-}\Big), 
\hspace{4mm} \alpha=1,\dots,8, \label{uexp}
\end{eqnarray}
with 
\begin{eqnarray}
&&[a^I_{pq}(k^+), {a^J_{rs}}^{\dagger}(k'^+)]=
\delta^{IJ}\delta_{pr}\delta_{qs} 
\delta(k^+- k'^+), \\
&&\{ b^{\alpha}_{pq}(k^+), {b^{\beta}_{rs}}^{\dagger}(k'^+)\}=
\delta^{\alpha\beta}
\delta_{pr}\delta_{qs}\delta(k^+- k'^+).
\end{eqnarray}
An important simplification of the light-cone quantization is that 
the light-cone vacuum  is the Fock vacuum $\vert 0 \rangle$, defined by   
\begin{equation}
a^I_{pq}(k^+)\vert 0 \rangle =b^{\alpha}_{pq}(k^+)\vert 0 \rangle=0, 
\end{equation}
for all positive longitudinal momenta $k^+ > 0$. 
We therefore have $P^+\vert 0 \rangle= P^-\vert 0 \rangle=0$
if we formulate the theory in the continuum, since the 
(zero measure) point
$k^+ = 0$ may be neglected, and the issue of
``zero modes'' does not arise\footnote{In the continuum
formulation, subtleties
associated with the singular point $k^+=0$ still arise,
but may be handled in the context of certain ``cancelation
conditions'' at vanishing longitudinal momentum $k^+ \rightarrow 0$.
See \cite{abd,me}, and references therein.}
\footnote{If we discretize the momenta, however,
such that the $k^+$ integrations are replaced by {\em finite} sums,
then the point $k^+ = 0$ can no longer be ignored, and the
``zero mode'' problem must be addressed \cite{yam}. In some 
cases, neglecting zero modes is legitimate even after discretizing 
momenta, and we expect that to be the case here. 
For recent work, see \cite{zero}, and references therein.}.  

\begin{figure}[h!]
\begin{center}
\epsfig{file=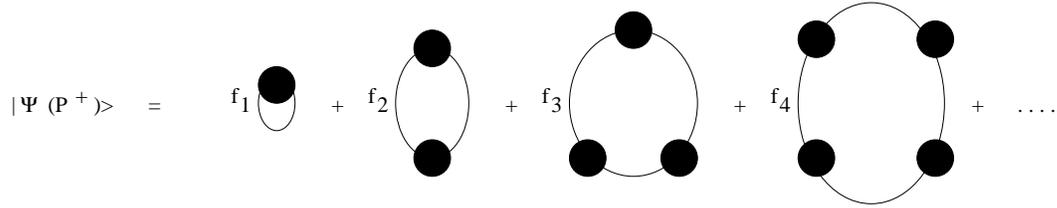,width=14cm}
\end{center}
\caption{{\small The single closed string expansion. The wavefunctions
$f_i$ depend on the light-cone momenta of the string bits, and summation
over all possible momenta is implied. The total momentum
of each closed string Fock state is $P^+$, and is
conserved in all interactions.} \label{pic2}}
\end{figure}

The ``charge-neutrality'' condition (first integral constraint
from (\ref{EQ4})) requires that all the colour indices must be 
contracted for physical states.
Thus the physical states are formed by colour traces of the 
boson and fermion creation operators 
${a^I}^{\dagger},{b^{\alpha}}^{\dagger}$
acting on the light-cone vacuum.  A single trace of these
creation operators may be identified
as a single closed string, where each operator
(or `parton'), carrying some longitudinal
momentum $k^+$,  represents a
 `bit' of the string. A product of traced operators
is then a multiple string state. A general superposition
of {\em single} closed strings  with total longitudinal
momentum $P^+$  takes the form 
\begin{eqnarray}
\lefteqn{ \vert \Psi(P^+) \rangle =
 \sum_{q=1}^{\infty} \int_0^{\infty} dk_1^+ \cdots dk_q^+ \hspace{1mm}
 \delta ( k_1^+ + \cdots + k_q^+ - P^+) \times } & & \nonumber \\
& & f_{\alpha_1 \cdots \alpha_q} (k_1^+,\dots, k_q^+)
\frac{1}{\sqrt{N^q}} \mbox{tr}[
 \Gamma_{\alpha_1}^{\dagger}(k_1^+) \cdots 
\Gamma_{\alpha_q}^{\dagger}(k_q^+) ] \vert 
0 \rangle , \label{fockexp1}
\end{eqnarray}
where the repeated indices $\alpha_i$ are summed over the 
eight boson and eight fermion degrees of freedom
such that $\Gamma_{\alpha_i}$ may represent any
boson operator $a^I$, or fermion operator $b^{\alpha}$.
The wavefunctions $f_{\alpha_1 \dots \alpha_q}$ are 
normalized such that  the
orthonormality condition $\langle \Psi(P^+) \vert 
\Psi(Q^+) \rangle
= \delta (P^+ - Q^+)$ holds. A simple diagramatic representation
of the expansion (\ref{fockexp1}) is shown in Figure \ref{pic2}.
Each solid disk represents an $N \times N$ bosonic or
fermionic  matrix operator $\Gamma_{\alpha_i}^{\dagger}$, 
and the lines connecting them
denote contraction of the matrix indices in the trace. These disks are 
the `string bits' which each carry some fraction of the total light-cone
momentum $P^+$. States which are dominated by an infinite number 
of such partons are evidently candidates for our string states.
We will elaborate on this remark in Section \ref{branes}.
\begin{figure}[h!]
\begin{center}
\epsfig{file=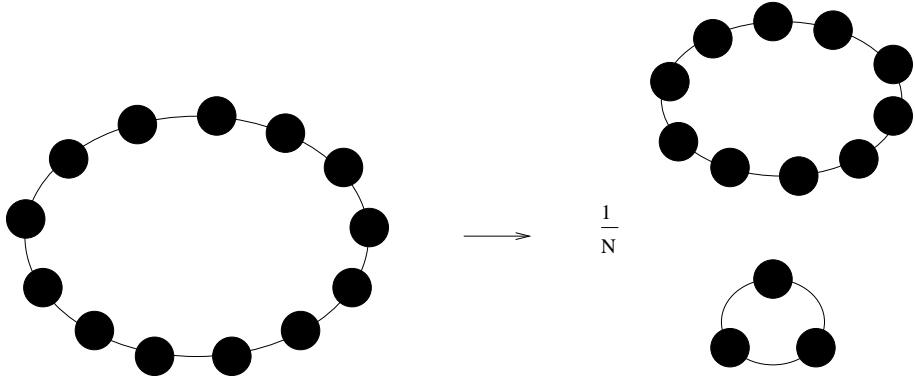,height=5cm}
\end{center}
\caption{{\small Splitting interactions, such as the one illustrated
above, are suppressed by the
factor $1/N$, which acts as a string coupling constant.
One would expect the absence of such processes in the large $N$
limit, but the existence of a double scaling limit 
may give rise to an effectively non-zero string coupling even
in the $N \rightarrow \infty$ limit.}  
 \label{pic3}}
\end{figure}
States involving multiple closed string states 
correspond in this formalism to a product
of two or more traces in the Fock space representation.
A Fock state with two closed strings, for example,
would have the general form
\begin{equation}
\frac{1}{\sqrt{N^q}} \mbox{tr}[
 \Gamma_{\alpha_1}^{\dagger}(k_1^+) \cdots 
\Gamma_{\alpha_q}^{\dagger}(k_q^+) ] \cdot 
\frac{1}{\sqrt{N^s}} \mbox{tr}[
 \Gamma_{\beta_1}^{\dagger}({\tilde k}_1^+) \cdots 
\Gamma_{\beta_s}^{\dagger}({\tilde k}_s^+) ] \vert 0 \rangle,
\end{equation}
where $\sum_{i=1}^{q} k_i^+ + \sum_{j=1}^{s} {\tilde k}_j^+ = P^+$.
Finally, it should be stressed that the space of states generated
by the single closed string states (i.e. all states which may be written
as in (\ref{fockexp1})) forms an invariant subspace of the light-cone 
Hamiltonian in the large $N$ limit. The reasoning here is that
the splitting or joining interactions 
(see, for example, Figure \ref{pic3}) involving 
multiple string states
is suppressed by a factor $1/N$, which represents (in this
scenario) the string coupling
constant. This is the conventional interpretation of large $N$ Yang-Mills:
all non-planar diagrams are suppressed.
However, we shall discuss in Section \ref{doublescaling} how
a double scaling limit involving $N$, and an ultraviolet cutoff
$K$, enables one to have an effectively non-zero string coupling
even in the large $N$ limit. But first we need to introduce
the concept of Discretized Light-Cone Quantization in order
to define the integer $K$.

\section{Discretized Light-Cone Quantization (DLCQ)}
\label{dlcq}
If we substitute the mode expansions (\ref{Xexp}) and
(\ref{uexp}) for the 
bosonic and fermion fields into expressions (\ref{P+}) 
and (\ref{P-}),
we may explicitly derive the quantized light-cone momentum and
Hamiltonian operators $P^{\pm}$ in terms of the  
(momentum dependent) creation and 
annihilation operators $a^I,{a^I}^{\dagger}$, (bosons) and $b^{\alpha}, 
{b^{\alpha}}^{\dagger}$ (fermions)\cite{me2}.     
One may then extract boundstates and masses by solving the
eigen-equation
\begin{equation}
           2P^+ P^- \vert \Psi \rangle = M^2 \vert \Psi \rangle,
\label{mass-shell}
\end{equation}
where $\vert \Psi \rangle$ is some appropriate superposition 
of single and multiple closed string Fock states.
One can show that $P^+$ commutes with the Hamiltonian $P^-$
(i.e. it is conserved in all interactions), and is
already diagonal on the space of closed string Fock states.
The problem, therefore, is to diagonalize the light-cone
Hamiltonian $P^-$ with respect to the given Fock basis. 
If we substitute the most general closed string
expansion (involving single and multiple strings) for
$\vert \Psi \rangle$ into the eigen-equation (\ref{mass-shell}), 
we obtain an infinitely-coupled set of integral equations
relating wavefunctions from different Fock sectors.   
Finding  analytical solutions to these integral equations is
in general a formidable task, and it is here that 
the DLCQ method has proven to be 
extremely useful in extracting numerical solutions.
 In
the context of supersymmetric field theories, additional
simplifications can be made by noting that in certain
cases we may write $P^+P^- \sim (Q^+Q^-)^2$
in terms of the light-cone supercharges $Q^+,Q^-$, and so the 
eigen-problem in this case is reduced to diagonalizing (via DLCQ)
the square root  $Q^+Q^-$ \cite{sakai,kleb}. Application
of this technique to calculate  boundstates in Matrix String
Theory is the subject of current investigation\cite{me2}.  

Let us now briefly review the DLCQ method in the context of
Matrix String Theory (for a more detailed treatment, see \cite{pauli}).
The essential idea is surprisingly
simple: we  discretize the light-cone momenta of constituent
partons, or string bits, keeping in mind that the total 
light-cone momentum $P^+$ is always conserved. 
In practice, this means we introduce
an (ideally large) positive integer $K$ such that 
$P^+/K$ defines the smallest unit of momentum. 
The light-cone momentum $k^+$ of any constituent parton 
is then some positive integer multiple of this smallest unit:
\begin{equation}
             k^+ = \frac{n}{K} P^+, \hspace{5mm} n=1,2,\dots.
\end{equation}
Of course, we recover the continuum formulation
in the limit $K \rightarrow \infty$.
 Note that for a Fock state with $q$ partons,
with each parton carrying momentum $k_i^+$, this prescription gives
\begin{equation}
 k_1^+ + \cdots + k_q^+ = P^+; \hspace{8mm} k_i^+ = \frac{n_i}{K}P^+, 
\hspace{5mm} (i=1,\dots,q),
\label{cond1}   
\end{equation} 
where the integers $n_i$ lie in the range $1 \leq n_i \leq K$, and satisfy
the constraint 
\begin{equation}
n_1 + \cdots + n_q = K.
\label{cond2}
\end{equation}
Evidently, if $K$ is fixed and finite, then (\ref{cond2})
is satisfied in only a finite number of ways, and so 
the space of Fock states is also finitely enumerated.
In this case, the Hamiltonian is just a finite matrix, which,
in principle, can always be diagonalized by some numerical routine.
All of this depends on the crucial assumption that one may neglect
the ``zero modes''; i.e. $n_i > 0$.  
 In the
continuum formulation, we may certainly
assume $k^+ > 0$, since the point $k^+ = 0$ is a zero measure 
set, and cannot affect the evaluation of an integral.
However, this
is no longer the case in the discretized theory \cite{zero}. 
The crucial issue 
here is that for  finite $K$, integrals 
over light-cone momenta are 
replaced by finite sums, and so if there are any contributions
arising from integrations around singularities 
at 
$k^+ = 0^+$ in the continuum theory, then they can only
appear in the discretized version of the same theory via
the zero mode $k^+ = 0$. Even so, such zero modes may at 
best 
only provide a `mean field' picture of the true
dynamics at vanishingly small $k^+$ 
(since $K$ is finite) and so we are not
guaranteed of a faithful representation of the theory
even after we introduce these zero mode degrees of freedom.

One way to avoid the
issue of zero modes in the DLCQ formulation of Matrix
String Theory is to 
always assume the
continuum limit $K \rightarrow \infty$ in all expressions involving
$K$.    
The quantity $1/K$ is then to be interpreted
as vanishingly small, and so cannot represent some finite 
cutoff in the theory. Rather, any singularities that arise
at vanishing $k^+$ are to be taken care of by introducing a set of
cancelation conditions as in \cite{abd,me}. 

Admittedly, this formal approach is not always useful, 
since in most practical implementations of DLCQ one
works with finite values of $K$, and the continuum limit 
$K \rightarrow \infty$ is obtained by performing
a suitable extrapolation.
However, it should be emphasized that neglecting the zero 
mode in the discretized version of a number
of two dimensional theories
does not affect the spectrum of massive boundstates,
and so, in this case, the strategy of working with finite values of $K$,
and then extrapolating to the continuum limit $K \rightarrow \infty$
proves to be a remarkably effective way of 
numerically extracting boundstates
and mass spectra.
Perhaps the best candidates for such special theories
are two dimensional supersymmetric models with unbroken
supersymmetry. Matrix String Theory is therefore expected
to admit a numerical boundstate analysis via the DLCQ method
with finite $K$. 
 
\section{D-Partons, Wee-Partons, and Strings}
\label{branes}
Consider a Fock state representing a single closed string
of $q$ partons with total light-cone momentum $P^+ = \sum_{i=1}^q k_i^+$:
\begin{equation}
\frac{1}{\sqrt{N^q}} \mbox{tr}[
 \Gamma_{\alpha_1}^{\dagger}(k_1^+) \cdots 
\Gamma_{\alpha_q}^{\dagger}(k_q^+) ] \vert 
0 \rangle \label{fsw}
\end{equation}
where $\Gamma_{\alpha_i}^{\dagger}(k_i^+)$ is a creation operator 
for a fermion or boson carrying light-cone momentum $k_i^+$.
If we perform DLCQ, then the light-cone momenta must be integer
multiples of the smallest unit of momentum, $P^+/K$, and may
be identified with the $q$ integers $n_1,\dots,n_q$. These
integers need only satisfy the constraints
\begin{eqnarray}
& & n_i \geq 1, \hspace{4mm} i=1,\dots,q  \\
& & n_1 + \cdots + n_q = K \label{sumK},
\end{eqnarray}
For given (finite) values of $K$,  enumerating all such integer 
sets $(n_1,\dots,n_q)$  satisfying the above constraints is,
in principle, accomplished by a suitable computer algorithm,
although the processing time would increase at least exponentially
as $K$ is increased. What will be of interest to us are the general
physical properties that may be ascribed to such integer solution
sets in the continuum limit $K \rightarrow \infty$.

\begin{figure}[h!]
\begin{center}
\epsfig{file=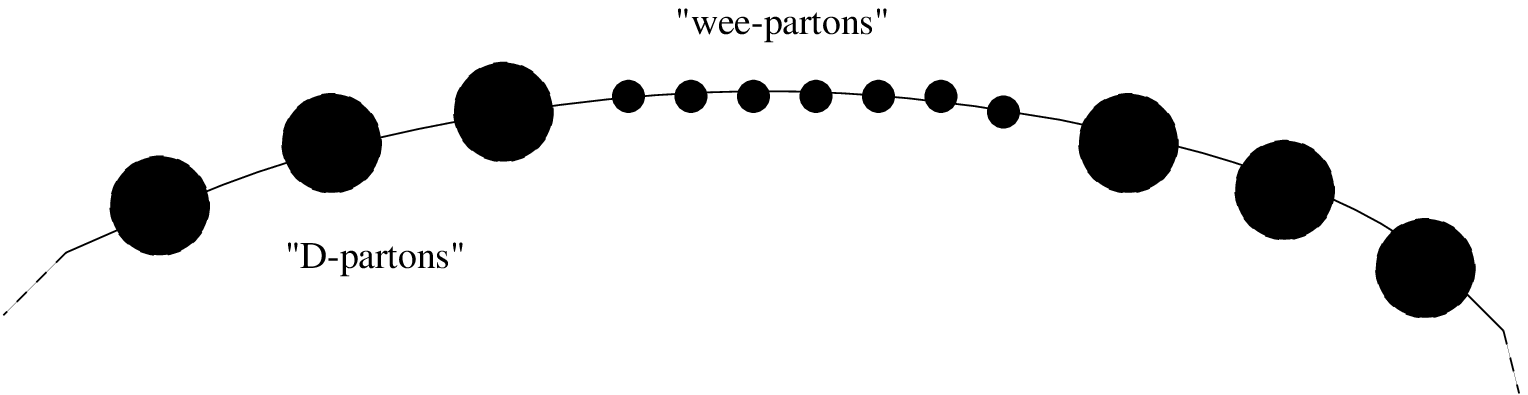,width=10cm}
\end{center}
\caption{{\small ``D-partons'' and ``wee-partons'' will be distinguished 
diagrammatically according to size. }  
 \label{pic4}}
\end{figure}

\begin{figure}[h!]
\begin{center}
\epsfig{file=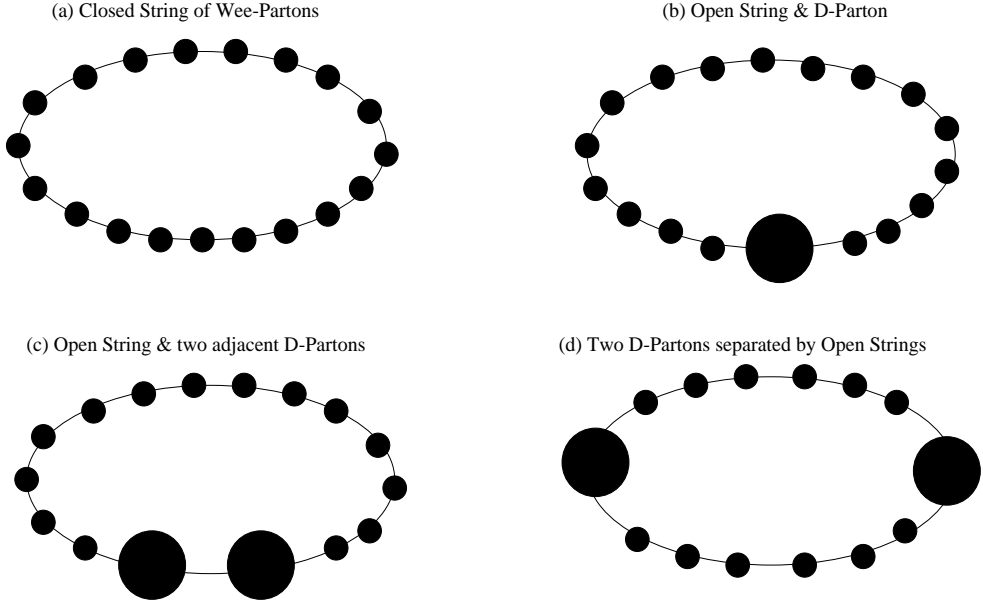, width=13cm, height=8cm}
\end{center}
\caption{{\small (a) A closed string made up from an infinite
number of wee-partons. (b) A ``long'' open string of wee-partons
ending on a single D-parton. (c) A long string of wee-partons ending
on different D-partons. (d) Two D-partons separated by two
long strings of wee-partons. }  
 \label{pic5}}
\end{figure}

The first case we consider is when the total number of
string bits, $q$, is finite. Since $K \rightarrow \infty$,
we must have at least one of the integers in the constraint
(\ref{sumK}) tend to infinity as well. We will call the corresponding 
partons in this case ``D-partons''; all remaining
partons will be called ``wee-partons'': 
\begin{eqnarray}
\mbox{``D-Partons''} & : & 0 < \lim_{K \rightarrow \infty} \frac{n_i}{K} \leq 1
\hspace{4mm} \mbox{for some $i \in \{1,\dots,q \}$;} \\
\mbox{``Wee-Partons''} & : &  \lim_{K \rightarrow \infty} \frac{n_i}{K} = 0,
\hspace{4mm} \mbox{for some $i \in \{1,\dots,q \}$.} 
\end{eqnarray}
Of course, this is just a formal way of saying that D-partons each have 
(positive) non-zero light-cone momentum, while the momentum 
carried by each wee-parton is  
vanishingly small. 
Pictorially (see Figure \ref{pic4}), 
we will distinguish a D-parton from a wee-parton 
by size.
Now consider what happens if we allow the total number of partons, $q$,
tend to infinity.
It is now possible to have an infinite number of wee-partons, even
though as a whole they may carry only a finite fraction of the total
light-cone momentum $P^+$.  
For example, let us consider the 
Fock state (\ref{fsw})  where
each parton carries the smallest possible unit of momentum 
($P^+/K$). In this case,
$n_i = 1$ for all $i=1,\dots,q$, and constraint (\ref{sumK}) implies
that the total number of such partons is $K$ (i.e. $q=K$).
As we let $K \rightarrow \infty$, we end up with a {\em closed string}
consisting of an infinite number of wee-partons (Figure \ref{pic5}(a)).

Now let us consider a state with a single D-parton. This may be
accomplished by making the assignments
\begin{equation} 
n_i = \left\{ \begin{array}{cl}
                  1 & \mbox{for $i=1,\dots,q-1;$} \\
                  K/2 & \mbox{for $i=q$}
              \end{array} \right.
\end{equation}
where $K$ is an even integer here, and $q=K/2+1$. The total
number of wee-partons is thus $K/2$, which tends to
infinity in the $K \rightarrow \infty$ limit. Half of
the total light-cone momentum is carried by this string
of wee-partons, while
the single D-parton 
carries the remaining half. This state
resembles an {\em open string} of wee-partons 
with both ends fixed to
the same D-parton (Figure \ref{pic5}(b)).
  
For an example of an open
string of wee-partons ending on different D-partons
(Figure \ref{pic5}(c)), we 
make the assignments
\begin{equation} 
n_i = \left\{ \begin{array}{cl}
                  1 & \mbox{for $i=1,\dots,q-2;$} \\
                  K/4 & \mbox{for $i=q-1$ and $i=q$}
              \end{array} \right.
\end{equation}
where $q=K/2+2$, and $K$ is divisible by four. In
the continuum limit $K \rightarrow \infty$, we have
two adjacent D-partons carrying half the total light-cone
momentum, while the ``long string'' of wee-partons (there being an 
infinite number of them) carries the other half. We will
use the phrase ``short string'' of wee-partons if there are 
only a finite number of them.
                
As a final example, we may construct two open
strings of wee-partons ending on two D-partons (Figure \ref{pic5}(d))
by making the assignments 
\begin{equation} 
n_i = \left\{ \begin{array}{cl}
                  1 & \mbox{for $i=1,\dots,q-K/2,$
            and $i=q-K/2+2,\dots,q-1;$} \\
                  K/4 & \mbox{for $i=q-K/2+1$ and $i=q$}
              \end{array} \right.
\end{equation}
where $q=K/2+2$ and $K$ is divisible by four. In this case,
the two D-partons are separated by long strings of wee-partons,
and collectively share half of the total light-cone momentum.

Evidently, these examples only provide a glimpse at the 
 the totality of all possible configurations, and whether 
a particular configuration dominates in a low energy 
boundstate or not depends on the dynamical properties of the 
light-cone Hamiltonian. Actually, in a recent
study of QCD coupled to adjoint fermions,
low energy boundstates made from only
wee partons were found to exist\cite{pin2}. 
In any case, one can see from these simple
observations that the existence of such
``long strings'' of wee-partons,
along with the D-partons, naturally incorporates
the dynamics of open and closed strings, together with D-particle-like
objects on which these open strings may end. However, the theory
we have constructed may exhibit structures which do not
admit an obvious classification from such string and/or 
D-particle-like configurations, and more detailed studies of
the Hamiltonian will be required if we wish to
investigate the structure
of the low energy boundstates in the theory.

\section{The Large $N$ Limit and Double Scaling}
\label{doublescaling}
In a fully interacting string theory, one has
``splitting'' and ``joining'' interactions 
which can change the number of  strings.
In the present context, if we take the large $N$ limit
while keeping $g^2 N$ fixed,\footnote{The constant
$g^2 N$ has the dimension of mass squared, which acts as the string
tension.} then by keeping track
of normalizations one can easily show that such
interactions are suppressed by a factor $1/N$ (Figure \ref{pic3}). 
However, in the DLCQ formulation, we have an additional
parameter $K$, which is also sent to infinity to
recover the continuum, and so there is now a possibility
of double scaling between these two diverging integers.
In particular, we would like to know whether  
one can still have an effectively non-zero 
string coupling constant even in the limit 
$N \rightarrow \infty$.

To investigate these ideas further,
we begin by considering the mass term operator in any 
massive two dimensional
bosonic matrix field theory (on the light-cone): 
\begin{equation}
    P^-_{mass} =  
\frac{1}{2}m^2 \int_0^{\infty} \frac{dk^+}{k^+} a_{ij}^{\dagger}(k^+) 
   a_{ij}(k^+). 
\end{equation}
Such an operator acts on the closed string of $q$ partons 
$\vert \Psi_q \rangle 
=\frac{1}{\sqrt{N^q}} \mbox{tr} \left[ a^{\dagger}(k_1^+) \cdots
      a^{\dagger}(k_q^+) \right] \vert 0 \rangle$ as
follows:
\begin{equation}
  P^-_{mass} \cdot \vert \Psi_q \rangle  
= \frac{1}{2} m^2 \left(\frac{1}{k_1^+} + \cdots + \frac{1}{k_q^+} \right)
\cdot 
\vert  \Psi_q \rangle,
\end{equation} 
where $k_1^+ + \cdots + k_q^+ = P^+$. In the DLCQ formulation,
the last identity becomes
\begin{equation}
 P^-_{mass} \cdot \vert \Psi_q \rangle  
= \frac{K}{P^+} 
\times \frac{1}{2}m^2 \left(\frac{1}{n_1} + \cdots + \frac{1}{n_q} \right)
\cdot \vert  \Psi_q \rangle,
\end{equation}        
 where the integers $n_i \in \{1,\dots,K\}$ are related to the original
light-cone momenta $k_i^+$ in the usual way: $k_i^+ = \frac{n_i}{K}P^+$,
and $n_1 + \cdots + n_q = K$. If $\vert  \Psi_q \rangle$ consists
only of D-partons (i.e. if $n_i/K$ is non-zero in the continuum
limit $K \rightarrow \infty$) then 
$\sum_{i=1}^q\frac{1}{n_i} \sim \frac{1}{K}$, and therefore
the mass operator is simply multiplication
by a finite constant which is indepnedent
of $K$: $P^-_{mass} \vert  \Psi_q \rangle
\sim \mbox{const.} \times \vert  \Psi_q \rangle$ as 
  $K \rightarrow \infty$. 

Now consider the case where $n_i=1$ for some finite number
of partons. i.e. we have a finite number of wee-partons.
In this case,  
$P^-_{mass} \vert  \Psi_q \rangle
\sim K \times \vert  \Psi_q \rangle$ as 
$K \rightarrow \infty$, and we therefore have 
a different scaling behaviour for the mass operator.

Finally, if we set all the integers $n_i$ to unity, so that
we end up with a closed string of (only) wee-partons,
then  $P^-_{mass}  \vert  \Psi_q \rangle
\sim K^2  \times \vert  \Psi_q \rangle$ as 
$K \rightarrow \infty$, and we thus arrive at another
scaling behaviour for the mass operator
when the number of wee-partons is allowed to
grow to infinity.  

In summary, we have observed in a simple case that 
operators may scale differently with respect to  $K$
depending on the Fock state configuration, and in particular,
long strings of wee-partons give rise to the largest scaling
exponent for $K$.

For Matrix String Theory (which has no explicit mass terms), the scaling
behaviour is perhaps 
not immediately calculable without solving
the full Hamiltonian -- a task which is under current
investigation \cite{me2} -- and so at this stage
our presentation will have to be schematic. 
At any rate, we know the theory is supersymmetric, and
so, schematically, we may write\footnote{Central
charges in the theory will be addressed 
in the forthcoming work \cite{me2}.} 
$P^- \sim \{Q^-,Q^- \}$, where $Q^-$ is a light-cone 
supercharge operator (in fact, there are eight such operators
in all -- see Appendix \ref{lcsuper}). The supercharge $Q^-$ contains
only cubic interactions\footnote{For more details, the reader
is referred to \cite{sakai} for a related treatment of a light-cone
supersymmetric matrix model.},
and in the DLCQ formulation has the
(very) schematic representation 
\begin{equation}
Q^- \sim \sqrt{g^2 N} \times \frac{K}{\sqrt{N}}
 \times \sum_{n,l=1}^{\infty}
\left(\frac{1}{n} + \frac{1}{l} \right)
  \times 
\left(\Gamma^{\dagger}(l+n) \Gamma(l) \Gamma(n)
+ \Gamma^{\dagger}(n)\Gamma^{\dagger}(l) \Gamma (n+l)\right),
\end{equation} 
where the operators
$\Gamma,\Gamma^{\dagger}$ annihilate and create partons in a
closed string of partons. Note that $g^2 N$  is held  fixed as
we take the large $N$ limit. Splitting interactions -- such as the
one illustrated in Figure \ref{pic3} -- are generated by
the terms  $\Gamma^{\dagger} \Gamma \Gamma$, and introduce
an overall factor of $\frac{1}{\sqrt{N}}$, so the amplitude
for a single splitting interaction is roughly
\begin{equation}
\frac{K}{N} \times \frac{1}{n} \hspace{6mm}
\mbox{(splitting amplitude)  }
\label{amp}
\end{equation}
For a D-parton, the integer $n$ (corresponding to
 the light-cone momentum $\frac{n}{K}P^+$) must scale 
like $n \sim K$, and so the amplitude (\ref{amp}) is 
$1/N$, which vanishes in the large $N$ limit.
This suggests that Fock states of D-partons
do not give rise to interacting string dynamics  
in the large $N$ limit. This is the usual interpretation
of large $N$ Yang-Mills: all non-planar diagrams 
are suppressed -- which, in this case, is equivalent to restricting
to the Fock space of {\em single} closed loops of partons. 

Now consider a long closed string of 
wee-partons. To make things simple,
assume each parton has the smallest possible light-cone momentum, $P^+/K$,
so that we have a closed string of $K$ wee-partons. The integer
$n$ appearing in (\ref{amp}) is now equal to unity, and 
the splitting amplitude is now $K/N$. If all the wee-partons
are identical, there may be an additional factor of $K$ which
arises from cyclic symmetry, but for now
we restrict ourselves to the more general case.
It is now clear that with the amplitude $K/N$ one has the possibility
of taking a double scaling limit; i.e. we let 
$N \rightarrow \infty$ and $K \rightarrow \infty$ while keeping
the ratio 
\begin{equation}
            \frac{K}{N} \equiv  \mbox{``string coupling''}
\end{equation}
fixed. Although the analysis here is very naive, it is nevertheless
intriguing that $K$ must scale in proportion to $N$ if we wish
to incorporate interacting string dynamics. It is of course
well established that the limit $K \rightarrow \infty$ in DLCQ
effects
a decompactification of a (light-cone) space dimension, and
so identifying $K$ with $N$ in the present context
suggests that the large $N$
limit is indeed associated with a decompactification
of an additional space-like dimension. Related ideas have
been proposed in a recent work by Susskind \cite{suss1}.   

At any rate, the resulting theory can no longer be identified
with conventional Yang-Mills, since we have an additional 
parameter specified by the ratio $K/N$, which acts as an
effective string coupling. Of course, for small values of this ratio,
we recover conventional large $N$ Yang-Mills  weakly
coupled to a string-like theory.

\section{Discussion}
\label{end}
Our investigations have shown that there is scope for 
quantitative investigations of field theories that are
generalizations of conventional large $N$ Yang-Mills theory.
In particular, we suggest that Discretized Light-Cone Quantization
offers a promising approach to unify in a natural
framework open and closed string dynamics coupled to
conventional large $N$ Yang-Mills theory. 

A key ingredient in this proposal
is the identification of a double scaling limit involving
the U($N$) gauge group parameter $N$, and the DLCQ
harmonic resolution $K$, which are both sent to infinity.
We provided a crude argument as to why the ratio
$K/N$ must be constant for an interacting string theory,
and how such a ratio is related to the effective string
coupling constant. This observation appears to
be consistent with 
the recent suggestion that  Matrix
Theory for finite $N$ corresponds to the DLCQ of $M$-Theory,
with harmonic resolution $K=N$.

In any case, it is clear that further numerical studies
of two (and possibly higher) dimensional Yang-Mills
theories 
in the large $N$ limit would help clarify 
the physical consequences that follow from
taking this limit. As we have seen,
the emergence of an additional parameter via a double
scaling limit, and the proposed connection with conventional
string theory, suggests (rather ironically)
that we are perhaps
understanding the full implications of
Yang-Mills theory for the first time.

\medskip
\begin{large}
 {\bf Acknowledgments}
\end{large}

The authors would like to thank S.Pinsky
and A.Curtis for many stimulating discussions.
 
\appendix
\section{Appendix: Yang-Mills in Ten Dimensions}
\label{ymills10}
Let's start with ${\cal N}=1$ super Yang-Mills theory 
in 9+1 dimensions with gauge group U($N$), where
$\theta = 0$ in the instanton contribution:
\begin{equation}
S_{9+1}=\int d^{10}x \hspace{1mm} \mbox{tr} \Bigg
(-\frac{1}{4} F_{\mu\nu}F^{ \mu\nu}+\frac{{\rm i}}{2} 
\bar{\Psi}\Gamma^{\mu}D_{\mu}\Psi\Bigg) , 
\label{EQ1}
\end{equation} 
where 
\begin{eqnarray}
F_{\mu\nu}&=&\partial_{\mu}A_{\nu}-\partial_{\nu}A_{\mu}
+{\rm i}g[A_{\mu},  A_{\nu}] , \\
D_{\mu}\Psi &=& \partial_{\mu}\Psi+{\rm i}g[A_\mu, \Psi],  
\end{eqnarray} 
and $\mu,\nu = 0,\dots,9$.
The Majorana spinor  $\Psi$ 
transforms in the adjoint representation of  U($N$). 
The (flat) space-time metric
$g_{\mu \nu}$ has signature $(+,-,\dots,-)$, and we adopt 
the normalization $\mbox{tr}(T^aT^b) = \delta^{a b}$ for
the generators of the U($N$) gauge group.

In order to realize the ten dimensional Dirac algebra  
$\{\Gamma_\mu, \Gamma_\nu\}=2g_{\mu\nu}$ in terms
of Majorana matrices, 
we use as building blocks the 
reducible ${\bf 8}_s + {\bf 8}_c$ representation
of the spin(8) Clifford Algebra. In block form, we have 
\begin{equation}
\gamma^I=\left(\begin{array}{cc}
0 & \beta_I\\
\beta_I^T & 0 
\end{array}\right), \hspace{7mm} I=1,\dots,8,
\end{equation} 
where the $8 \times 8$ real matrices, $\beta_I$, satisfy
$\{\beta_I,\beta_J^T \} = 2\delta_{IJ}$. This automatically
ensures the spin(8) algebra $\{\gamma^I,\gamma^J \} = 2\delta^{IJ}$
for the $16 \times 16$ real-symmetric matrices $\gamma^I$.
An explicit representation for the $\beta_I$ algebra 
may be given in terms of a tensor product of Pauli matrices \cite{schwarz}. 
In the present context, we may choose a representation such that
a ninth matrix, $\gamma^9 = \gamma^1 \gamma^2 \cdots \gamma^8$, which
anti-commutes with the other eight $\gamma^I$'s, 
takes the explicit form   
\begin{equation} 
\gamma^9=\left(\begin{array}{cc}
{\bf 1}_{8} & 0\\
0 & -{\bf 1}_{8} \end{array}\right). \label{gamma9}
\end{equation}
We may now construct $32 \times 32$ pure imaginary
(or Majorana) matrices $\Gamma^\mu$ which realize
the Dirac algebra for the Lorentz group SO($9,1$): 
\begin{eqnarray}   
&& \Gamma^0=\sigma_2 \otimes {\bf 1}_{16}, \\
&& \Gamma^I={\rm i}\sigma_1 \otimes \gamma^I, \hspace{6mm} I=1,\dots,8;\\
&& \Gamma^9= {\rm i}\sigma_1 \otimes \gamma^9.
\end{eqnarray}
The Majorana spinor therefore has 32 real components, and
since it transforms in the adjoint 
representation of U($N$),
each of these components may be viewed as an $N \times N$ 
Hermitian matrix. 

An additional matrix $\Gamma_{11}= 
\Gamma^0 \cdots \Gamma^9$, which is
equal to $\sigma_3\otimes {\bf 1}_{16}$ in the representation
specified by (\ref{gamma9}), 
is easily seen to anti-commute with all other gamma matrices, and
satisfies  $(\Gamma_{11})^2 = 1$. It is also real, and so 
the  Majorana spinor field $\Psi$ admits a chiral decomposition
via the projection operators $\Lambda_{\pm} \equiv 
\frac{1}{2}(1 \pm \Gamma_{11})$:
\begin{equation}
 \Psi = \Psi_+ + \Psi_-, \hspace{5mm}  \Psi_{\pm} =  \Lambda_{\pm} \Psi.
\end{equation}
We will therefore consider only spinors 
with positive chirality $\Gamma_{11} \Psi = +\Psi$ (Majorana-Weyl):
\begin{equation}
\Psi= 2^{1/4} { \psi \choose 0}, \label{spin16} 
\end{equation}         
where $\psi$ is a sixteen component real spinor, and the 
numerical factor $2^{1/4}$ is introduced for later convenience. 

Since $\gamma^9$ anti-commutes with the other eight $\gamma^I$'s,
and satisfies $(\gamma^9)^2 = 1$, we may construct further
projection operators $P_R \equiv \frac{1}{2}(1+\gamma^9)$ and
$P_L \equiv \frac{1}{2}(1-\gamma^9)$ which project out, respectively, the 
right-moving and left-moving components of the sixteen component
spinor $\psi$ defined in (\ref{spin16}):
\begin{equation}
 \psi = \psi_R + \psi_L, \hspace{5mm}  \psi_R =  P_R \psi, \hspace{3mm}
\psi_L =  P_L \psi.
\end{equation}
This decomposition is particularly useful when working
with light-cone coordinates, since in the light-cone gauge
one can express the left-moving component $\psi_L$ in terms of
the right-moving component $\psi_R$ by virtue of the
fermion constraint equation. We will derive this result shortly.
In terms of the usual ten dimensional Minkowski
space-time coordinates, the light-cone coordinates are given by
\begin{eqnarray}
 x^+ & = & \frac{1}{\sqrt{2}}(x^0 + x^9), \hspace{10mm}
\mbox{``time coordinate''} \\
 x^- & = & \frac{1}{\sqrt{2}}(x^0 - x^9), \hspace{10mm}
\mbox{``longitudinal space coordinate''}  \\ 
 {\bf x}^{\perp} & = & (x^1,\dots,x^8).
\hspace{13mm} \mbox{``transverse coordinates''} 
\end{eqnarray}
Note that the `raising' and `lowering' of the $\pm$ indices
is given by the rule $x^{\pm} = x_{\mp}$, 
while $x^I = -x_I$ for $I=1,\dots,8$,
as usual. It is now a routine task to demonstrate that
the Yang-Mills action (\ref{EQ1}) for the positive
chirality spinor (\ref{spin16}) is equivalent to
\begin{eqnarray}
S_{9+1}^{LC} & = & \int dx^+ dx^- d{\bf x}^{\perp} \hspace{1mm}
 \mbox{tr} \left( \frac{1}{2}F_{+-}^2 + F_{+I}F_{-I} - \frac{1}{4}F_{IJ}^2
 \right. \nonumber \\
& & \hspace{20mm}
+ \hspace{1mm}
{\rm i} \psi_R^T D_+ \psi_R +   {\rm i}\psi_L^T D_- \psi_L +
     {\rm i}\sqrt{2}\psi_L^T \gamma^I D_I \psi_R \left. 
\frac{}{} \right),
\label{LCversion}
\end{eqnarray}
where the repeated indices $I,J$ are summed over $(1,\dots,8)$.
Some surprising simplifications follow if we now choose 
to work in the {\em light-cone gauge} $A^+ = A_- = 0$. In
this gauge $D_- \equiv \partial_-$, and so the (Euler-Lagrange)
equation of motion for the left-moving field $\psi_L$
is simply 
\begin{equation}
 \partial_- \psi_L = -\frac{1}{\sqrt{2}}\gamma^I D_I \psi_R, 
\label{fermioncon}
\end{equation}
which is evidently a non-dynamical constraint equation, since it
is independent of the light-cone time. We may therefore eliminate
any dependence on $\psi_L$ (representing unphysical 
degrees of freedom) in favour of $\psi_R$, which carries the
eight physical fermionic degrees of freedom in the theory.    
In addition, 
the equation of motion for the $A_+$ field yields
Gauss' law:
\begin{equation} 
\partial_{-}^2 A_{+}=\partial_{-}\partial_{I}A_{I}+gJ^+
\label{apluscon}
\end{equation} 
where $J^+={\rm i}[A_{I},\partial_{-}A_{I}]+2\psi_{R}^T\psi_{R}$, and
so the $A_+$ field may also be eliminated
to leave the eight bosonic degrees of freedom $A_I$, $I=1,\dots,8$.
Note that the eight fermionic degrees of freedom
exactly match the eight
bosonic degrees of freedom associated with the transverse
polarization of a ten dimensional gauge field, which is of
course consistent
with the supersymmetry. We should emphasize that
unlike the usual 
covariant formulation of Yang-Mills, the light-cone formulation
here permits one to remove {\em explicitly} 
any unphysical degrees of freedom in the 
Lagrangian (or Hamiltonian); there 
are no ghosts. 


\section{Light-Cone Supersymmetry}
\label{lcsuper}
The supercharges of ${\cal N}=8$ Matrix String Theory 
can be obtained by the dimensional 
reduction of the supercharge of ${\cal N}=1$ 
super Yang-Mills in ten dimension.
The time component of the reduced ten dimensional
supercurrent may be decomposed as follows: 
\begin{equation}
j^+=\frac{1-\gamma^9}{2}j^++\frac{1+\gamma^9}{2}j^+, 
\end{equation}
where 
\begin{eqnarray}
&&\frac{1-\gamma^9}{2}j^+=2^{\frac{5}{4}}
\partial_-X_I \gamma^I \theta_R, \\
&&\frac{1+\gamma^9}{2}j^+=2^{\frac{3}{4}} \partial_-A_+\theta_R + 
{\rm i} 2^{-\frac{1}{4}}g [X_I, X_J] \gamma^{IJ}\theta_R, 
\end{eqnarray}
and $\gamma_{IJ}=[\gamma_I, \gamma_J]/2$. 
After eliminating the non-dynamical variables and 
introducing the eight-component real spinor $u$, 
the supercharges of ${\cal N}=8$  Matrix String Theory
on the light-cone are given by ($\alpha =1,\dots,8$):
\begin{eqnarray}
&&Q_{\alpha}^+=\int dx^- 2^{\frac{5}{4}}
(\partial_- X_I \beta^T_I u_\alpha), \\
&&Q_{\alpha}^- =g 
\int dx^- \Bigg(2^{\frac{3}{4}} \partial_{-}^{-1}J^+ u_{\alpha} 
+{\rm i} 2^{-\frac{5}{4}}[X_I, X_J] 
(\beta_I\beta^T_J-\beta_J\beta^T_I)_{\alpha\beta} \cdot u_\beta \Bigg).  
\end{eqnarray}

\vfil

\end{document}